\documentclass[english,prb,floatfix,twocolumn]{revtex4-2}
\usepackage{ae,aecompl}
\usepackage[T1]{fontenc}
\usepackage[utf8]{inputenc}
\usepackage{babel}
\usepackage{float}
\usepackage{amsmath}
\usepackage{amssymb}
\usepackage{graphicx}
\usepackage{wasysym}
\usepackage{color}
\usepackage[unicode=true, bookmarks=true, bookmarksnumbered=false, bookmarksopen=false,
 breaklinks=false,pdfborder={0 0 1}, backref=false,colorlinks=true, linkcolor=magenta, citecolor=blue]
 {hyperref}
\begin{document}
\global\long\def\ket#1{\left|#1\right\rangle }%
\global\long\def\bra#1{\left\langle #1\right|}%
\global\long\def\braket#1#2{\langle#1|#2\rangle}%

\title{Anomalous correlation-induced dynamical phase transitions}
\author{Niaz Ali Khan}
\email{niazkhan@zjnu.edu.cn}
\affiliation{Department of Physics, Zhejiang Normal University, Jinhua 321004, P. R. China}

\author{Pei Wang}
\affiliation{Department of Physics, Zhejiang Normal University, Jinhua 321004, P. R. China}

\author{Munsif Jan}
\email{mjansafi@zjnu.edu.cn}
\affiliation{Department of Physics, Zhejiang Normal University, Jinhua 321004,
P. R. China}
\author{Gao Xianlong}
\email{gaoxl@zjnu.edu.cn}
\affiliation{Department of Physics, Zhejiang Normal University, Jinhua 321004, P. R. China}
\begin{abstract}
\section*{Abstract}
The nonanalyticity of the Loschmidt echo at critical times in quantum quenched systems is termed as the dynamical quantum phase transition, extending the notion of quantum criticality to a nonequilibrium scenario. In this paper, we establish a new paradigm of dynamical phase transitions driven by a sudden change in the internal spatial correlations of the disorder potential in a low-dimensional disordered system. The quench dynamics between prequenched pure and postquenched random system Hamiltonian reveals an anomalous quantum dynamical quantum phase transition triggered by an infinite disorder correlation in the modulation potential. The physical origin of the anomalous phenomenon is associated with the overlap between the two distinctly different extended states. Furthermore, we explore the quench dynamics between the prequenched random and postquenched pure system Hamiltonian. Interestingly, the quenched system undergoes dynamical quantum phase transitions for the prequench white-noise potential in the thermodynamic limit. In addition, the quench dynamics also shows a clear signature of the delocalization phase transition in the correlated Anderson model.
\end{abstract}
\maketitle

\section{Introduction}

Quantum phase transitions in nonequilibrium setting have become a topic of vivid interest in the field of condensed matter physics \citep{Heyl2015,Jurcevic2017,Flaschner2018,Mitra2018,Heyl2018,heyl2018detecting,Liu2018,Abdi2019,Ding2020,Yu2021,Peotta2021,Hamazaki2021,Wrzesniewski2022,Wong2022,Damme2022,Dborin2022,Zhou2023,Corps2023}.
Remarkably, the nonequilibrium phase transitions are driven by progressing
time which provides a new framework to explore the dynamical behavior
of time-evolving quantum systems \citep{Wrzesniewski2022,Wong2022,Jafari2021,Naji2022,Jafari2022,Niaz2023-DimerDQPT}.
In fact, the concepts of quantum criticalities in nonequilibrium setting
have been elegantly mapped to the dynamical quantum phase transitions
(DQPTs), where the singularities of Loschmidt echo identify the DQPTs
of quantum quenched systems \citep{Yang2017,Yin2018,Xu2021}. The
Loschmidt echo is a measure of the overlap between the reference and
time-evolved quantum states, which has been extensively studied both theoretically
\citep{Liu2018,Abdi2019,Ding2020,Yu2021,Peotta2021,Hamazaki2021,Wrzesniewski2022,Wong2022,Yang2017,Yin2018,Xu2021,Jafari2021,Jafari2022,Naji2022,Niaz2023-DimerDQPT}
and experimentally \citep{Jurcevic2017,Flaschner2018,Chen2020,Dag2021}. A paradigmatic
model showing DQPTs is the Aubry--André model after a quench
of the strength of incommensurate potential \citep{Yang2017,Xu2021}.
In addition, the nonequilibrium dynamics of the Anderson model after
a quench of the disorder strength has also been explored \citep{Yin2018}.
The concept of dynamical phase transitions may also be characterized
by entanglement echo \citep{Poyhonen2021,Nicola2021,Nicola2022} (the overlap of the initial and its time-evolved entanglement Hamiltonian ground states) of
the subsystems embedded in a larger quantum systems.
Moreover, the DQPTs can be probed by measuring
the nonequilibrium order parameter in the Lipkin-Meshkov-Glick model
with a quenched transverse field \citep{Xu2020}. 

Anderson localization is a quantum phase transition driven by the
uncorrelated disorder strength under certain conditions, as laid down
by the seminal work of Anderson \citep{Anderson1958}. In the context
of tight-binding, all eigenstates in noninteracting low-dimensional
systems are localized by an infinitesimal amount of disorder in the
thermodynamic limit \citep{greiner2002quantum}, whereas a three-dimensional system displays metal-insulator transition at critical disorder strength with a mobility edge separating extended and localized states \citep{Thouless1979,PhysRevLett.101.255702,Lagendijk2009,jendrzejewski2012three,Niaz2021}.

Correlations in disorder potential are known to lead to the quantum phase
transition in the noninteracting low-dimensional correlated disordered
system \citep{Moura1998,Petersen2013,Pires2019,Niaz2020,Paschen2021,Dikopoltsev2022}. Remarkably,
the correlated Anderson model displaying metal-insulator transition
at critical correlation exponent, $\alpha=2$, with a mobility edge
demarcating extended and localized states \citep{Moura1998}. The
transition was reaffirmed on the basis of strong anticorrelations
of the disordered potential in thermodynamic limit \citep{Petersen2013}.
With regard to the phase transition, Pires et al. \citep{Pires2019}
demonstrated that the delocalization phase transition may occur at $\alpha\sim1$ without a mobility edge in the perturbative regime.
It was found that the localization length diverge as $(1-\alpha)^{-1}$
in limit $\alpha\rightarrow1$ in the thermodynamic limit, confirmed by the analytical perturbative calculations \citep{Pires2019,Niaz2020}.

\begin{figure}[ht!]
\begin{centering}
\includegraphics[scale=0.74]{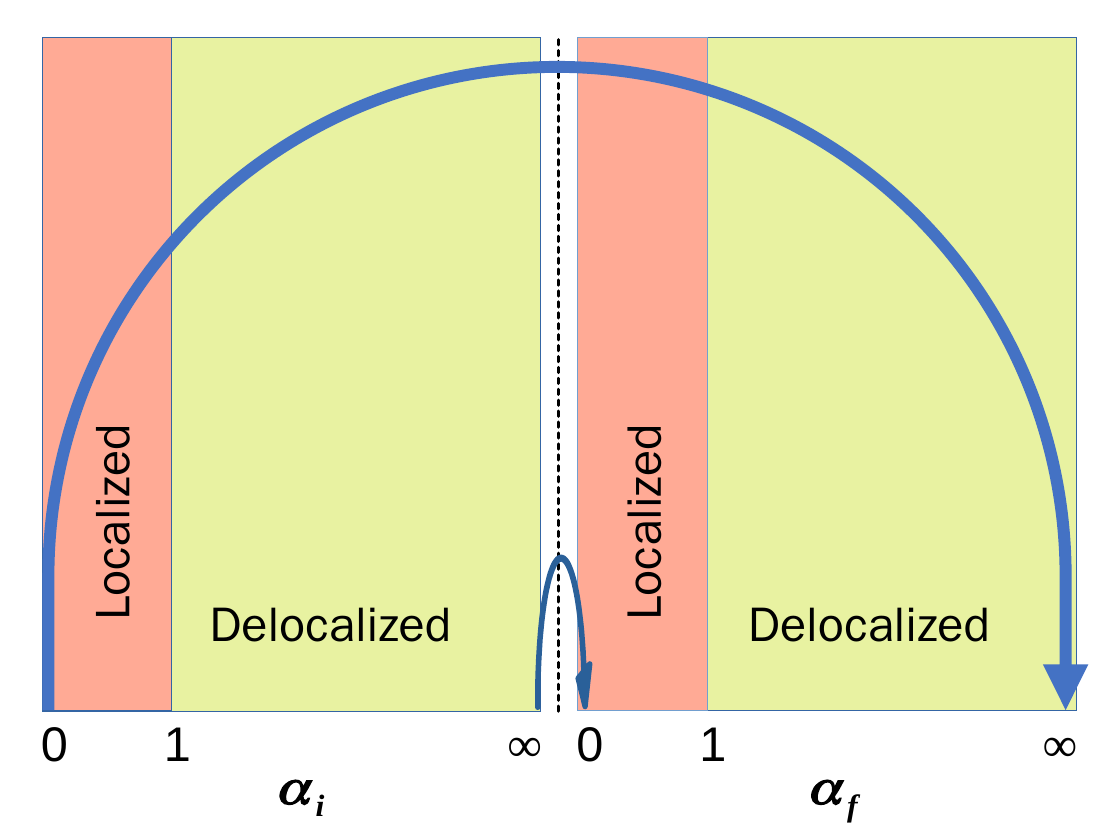}
\par\end{centering}
\caption{(Color online) A schematic representation of the quantum quench process under the correlated Anderson model. The parameters $\alpha_{i}$ and $\alpha_{f}$ control the prequench and postquench modulation potential strengths, respectively. Here, we show two extreme cases of quench dynamics (bold blue arrows), i.e., $\alpha_{i}=\infty (0)$, and $\alpha_{f}=0 (\infty)$.  An abrupt quench of the system variables triggers a dynamical phase transition in a lattice with $N$ sites. The black dashed line demarcates the prequench and postquench regimes.} \label{fig:QuenchingBD}
\end{figure}

Dynamical phase transition is a quantum critical phenomenon in nonequilibrium settings, characterized by the dynamical properties of quantum quenched systems. In this paper, we formulate a nonstationary dynamical evolution of noninteracting fermions with diagonal correlated random energies. The quantum quench dynamics is characterized by sudden changes in the internal correlations of the disorder potential. A schematic representation of quantum quench process for two limiting cases i.e., quench processes between states with (i) $\alpha_{i}=\infty$ (delocalized), $\alpha_{f}=0$, (localized), and (ii) $\alpha_{i}=0$ (localized), $\alpha_{f}=\infty$ (delocalized), is illustrated in Fig.~\ref{fig:QuenchingBD}. We obtain a universal feature of the Loschmidt echo for an initially prepared pure and strongly correlated time-evolving state. In this scenario, the Loschmidt echo becomes nonanalytically anomalous in critical times, signaling the correlation-induced DQPTs. However, conventionally, the Loschmidt amplitude always one for an initially ground and time-evolved extended states. On the other hand, the Loschmidt echo turns out to be size-dependent for an initially prepared localized and the time-evolving pure state. We further observe the delocalization transition in the correlated Anderson model from the perspective of the Loschmidt echo.

The structure of the paper is as follows. Sec.~\ref{sec:CorrelatedModel}
discusses the tight-binding model with the effect of diagonal
random energies. The randomness of the disorder potential is demonstrated
as a long-range correlated disorder under power-law spectral density. Sec.~\ref{sec:LoschmidtEcho} focuses on the properties of the Loschmidt echo in the perturbative regime for various correlation exponent. We discuss the dynamical signatures of the quantum phase transition characterized by the zeros of Loschmidt echo in critical times. Last section summarize our conclusions.
\section{The Correlated Anderson Model\label{sec:CorrelatedModel}}

\begin{figure}[ht!]
\begin{centering}
\includegraphics[scale=0.37]{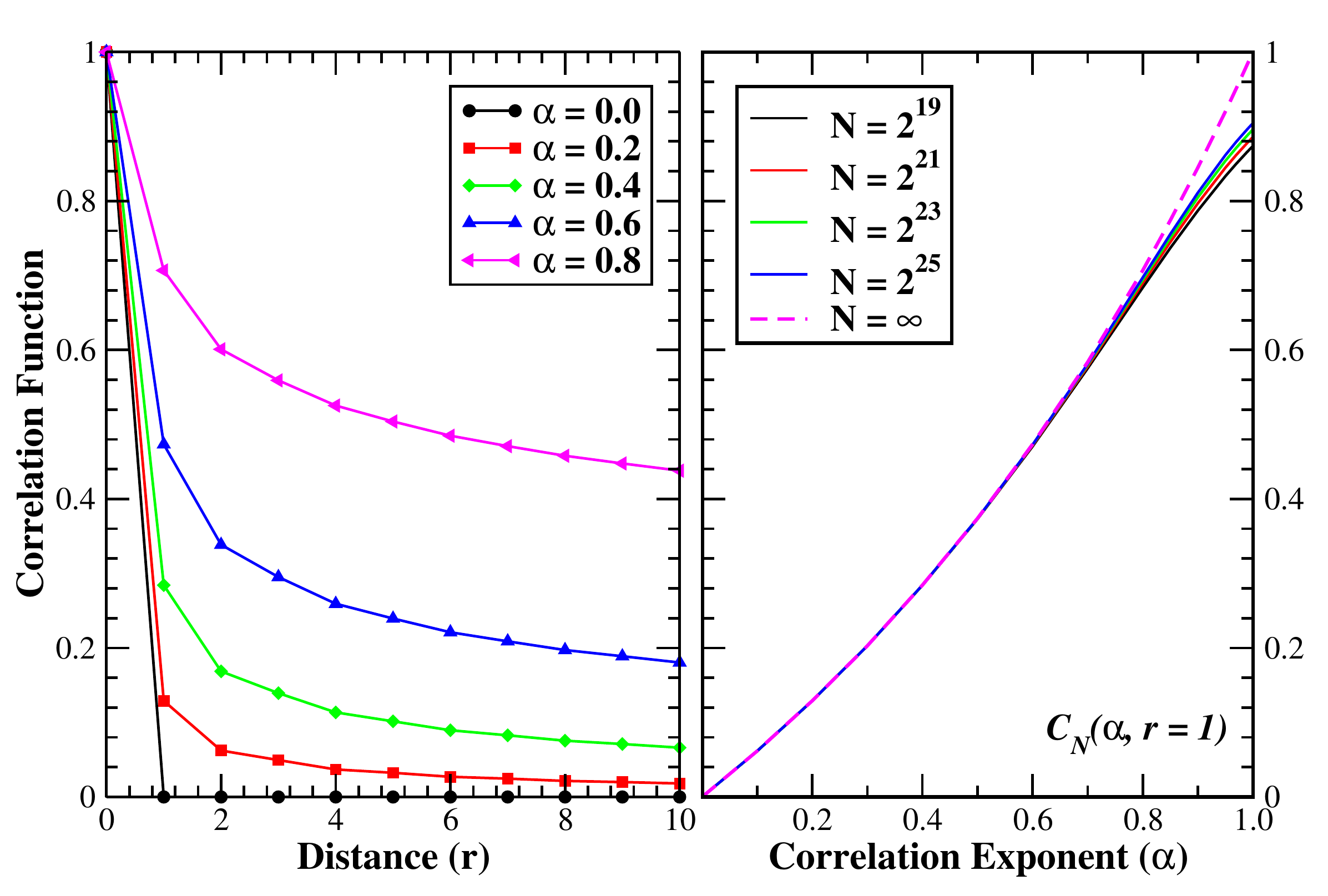}
\par\end{centering}
\caption{(Color online) Left panel: The two-point normalized correlation function
of the local disorder $\varepsilon_{n}$ in the thermodynamic limit.
The correlation function tends to unity in the limit $\alpha\sim1$.
Right panel: The correlation function as a function of $\alpha$ for
various system sizes at $r=1$. For finitely large system sizes, the
correlations converge very slowly towards the thermodynamic value
in the limit $\alpha\sim1$.\label{fig:AsymptoticCF}}
\end{figure}

Here, our model consists of noninteracting spinless electrons
in a disordered potential with long-range spatial correlations. The
Hamiltonian of our model has the general form \citep{Moura1998,Niaz2019,Niaz2023-EC1DCD},
\begin{equation}
\hat{\mathcal{H}}=-t\sum_{n=0}^{N-1}(c_{n}^{\dagger}c_{n+1}+c_{n+1}^{\dagger}c_{n})+\sum_{n=0}^{N-1}\varepsilon_{n}(\alpha)c_{n}^{\dagger}c_{n},\label{eq:1DHamiltonian}
\end{equation}
where $t$ denotes the transfer energy (hopping integrals) between the nearest neighboring
sites. For simplicity, $t=1,$ and all other energy scales are measured in unit of $t$.
In the second term of Hamiltonian, $\varepsilon_{n}$ represents the diagonal random energy of an
electron at the $n$-th site of the lattice of size $N$.
The randomness in the potential is demonstrated as a long-range spatially correlated disorder under spectral
density, $S(k)\sim k^{-\alpha},$ with $\alpha$ being the strength of correlation of the spectral density that controls the roughness of the potential landscapes. The disordered potential amplitude $\varepsilon_{n}(\alpha)$,
is given by \citep{Moura1998,Petersen2013,Pires2019,Niaz2020,Niaz2019,Niaz2022},
\begin{equation}
\varepsilon_{n}(\alpha)=\mathcal{A}_{\alpha}\sum_{k=1}^{N/2}\frac{1}{k^{\alpha/2}}\cos\left(\frac{2\pi k}{N}n+\phi_{k}\right),\label{eq:rp0}
\end{equation}
where $\mathcal{A}_{\alpha}$ is a normalization constant, imposing
unit variance of the local potential $(\sigma_{\varepsilon}^{2} = 1)$ with zero mean, and $\phi_{k}$
are the $N/2$ independent random phases which are uniformly distributed in
the interval {[}$0,\,2\pi]$. It is very important to emphasize that the
disorder distribution takes the following sinusoidal form of wavelength
$N$ with a vanishing noise,
\begin{align}
\varepsilon_{n}(\alpha\rightarrow\infty) & =\sqrt{2}\cos\left(\frac{2\pi}{N}n+\phi_{1}\right),\label{eq:SinusoidalForm}
\end{align}
in the limit of the infinite correlation of the disorder potential.
The disordered potential is a static cosine potential with a random phase, and its local value is dominated by a single term, $k=1$. As a consequence, the system exhibits metallic behavior due to the lack of effective disorder. In this limit, the spectral function of the correlated Anderson model shows identical behavior to the density of state in real space \citep{Niaz2019}. In the limit of $\alpha=0$, the system is insulating in nature, with all eigenstates localized. For a finite system, the normalized
correlation function, $\mathcal{C}_{N}(\alpha,r),$ of the disordered
potential can be formulated as \citep{Petersen2013,Pires2019},
\begin{align}
\mathcal{C}_{N}(\alpha,r) & \equiv\frac{\left\langle \varepsilon_{n}(\alpha)\varepsilon_{n+r}(\alpha)\right\rangle }{\left\langle \varepsilon_{n}^{2}(\alpha)\right\rangle }=\frac{\sum_{k=1}^{N/2}k^{-\alpha}\cos\frac{2\pi kr}{N}}{\sum_{k=1}^{N/2}k^{-\alpha}}.\label{eq:CorrelationFunction-Finite}
\end{align}
 In the thermodynamic limit,  the correlation function is linear for $\alpha=2$, convex for $\alpha>2$, and concave for $1 < \alpha<2$, near $\gamma\sim0$, whereas it becomes negative for
$\alpha>1$ near $\gamma\simeq1$, where $\gamma=2r/N$ is dimensionless
lattice distance with $\gamma \in [0,\,1]$ \citep{Petersen2013}. On the other hand, the normalized two-point correlation function of $\varepsilon_{n}$ exhibits the most remarkable characteristics for $\alpha\apprle1$.
The correlator is stationary in the thermodynamic limit, given by
\begin{align}
\mathcal{C}_{\infty}(\alpha,r) & ={}_{1}F_{2}\left(\frac{1-\alpha}{2};\frac{1}{2},1+\frac{1-\alpha}{2};-(\frac{\pi r}{2})^{2}\right),\label{eq:CorrelatorInfinity}
\end{align}
where $_{1}F_{2}(x)$ is a hypergeometric function. Its asymptotic
behavior decaying as $r^{\alpha-1}$ for long distances:
\begin{equation}
\mathcal{C}_{\infty}(\alpha,r)\propto r^{\alpha-1}
\end{equation}
The thermodynamic correlation function as a function of distance $r$ for various $\alpha$ is shown in Fig.~\ref{fig:AsymptoticCF} (left panel). The correlation function turns out to be the Kronecker-delta function, $\mathcal{C}_{\infty}(\alpha=0,r)=\delta_{r,0}$, in the limit $\alpha\rightarrow 0$, recovering the usual uncorrelated Anderson disorder. The
correlation increases with correlation exponent, tending to unity for
$\alpha\sim1$ in the thermodynamic limit as depicted in the right panel
of Fig.~\ref{fig:AsymptoticCF}. However, one can clearly see a very
slow convergence of the correlation at $r=1$ towards the thermodynamic
limit, especially for $\alpha\sim1$. 
Intuitively, at $r>1,$
the correlation functions converge to unity for $\alpha$ approaches
to one.
\section{The Loschmidt Echo\label{sec:LoschmidtEcho}}

A quantum quench process is an abrupt change in the Hamiltonian $\mathcal{\hat{\mathcal{H}}}(x)$
of a system, where $x$ denotes the strength of the quenched parameter.
At time $\tau=0$,  $\ket{\Psi(x)}$ is the initially prepared ground
state of the system with normalization condition $\braket{\Psi(x)}{\Psi(x)}=1$.
The Hamiltonian $\mathcal{\hat{\mathcal{H}}}(y)$ governs the time
evolution of the system over times $\tau>0$, reaching the unitary
evolving state,\citep{Yang2017,Yin2018,Xu2021}
\begin{equation}
\ket{\Psi(x,y,\tau)}=e^{-i\tau\mathcal{\hat{\mathcal{H}}}(y)}\ket{\Psi(x)}.
\end{equation}
A Loschmidt echo $\mathcal{L}(x,y,\tau)$ is the dynamical version
of the ground-state fidelity (return probability), defined as \citep{Yang2017,Yin2018,Xu2021},
\begin{equation}
\mathcal{L}(x,y,\tau)=\left|\braket{\Psi(x)}{\Psi(x,y,\tau)}\right|^{2}.\label{eq:LoschmidtEcho}
\end{equation}
It is a measure of the overlap between an initial
reference and the time-evolved state, plays a central role in characterizing
the DQPTs. The quantity $\braket{\Psi(x)}{\Psi(x,y,\tau)}$ is known
as the Loschmidt amplitude $\mathcal{G}(x,y,\tau)$ of the quenched
system. Phenomenologically, the quantum quenches trigger a time-evolving
state $\ket{\Psi(x,y,\tau)}$ under the postquench Hamiltonian $\mathcal{\hat{\mathcal{H}}}(y)$
from a reference state $\ket{\Psi(x)}$.
\begin{figure}[ht!]
\begin{centering}
\includegraphics[scale=0.34]{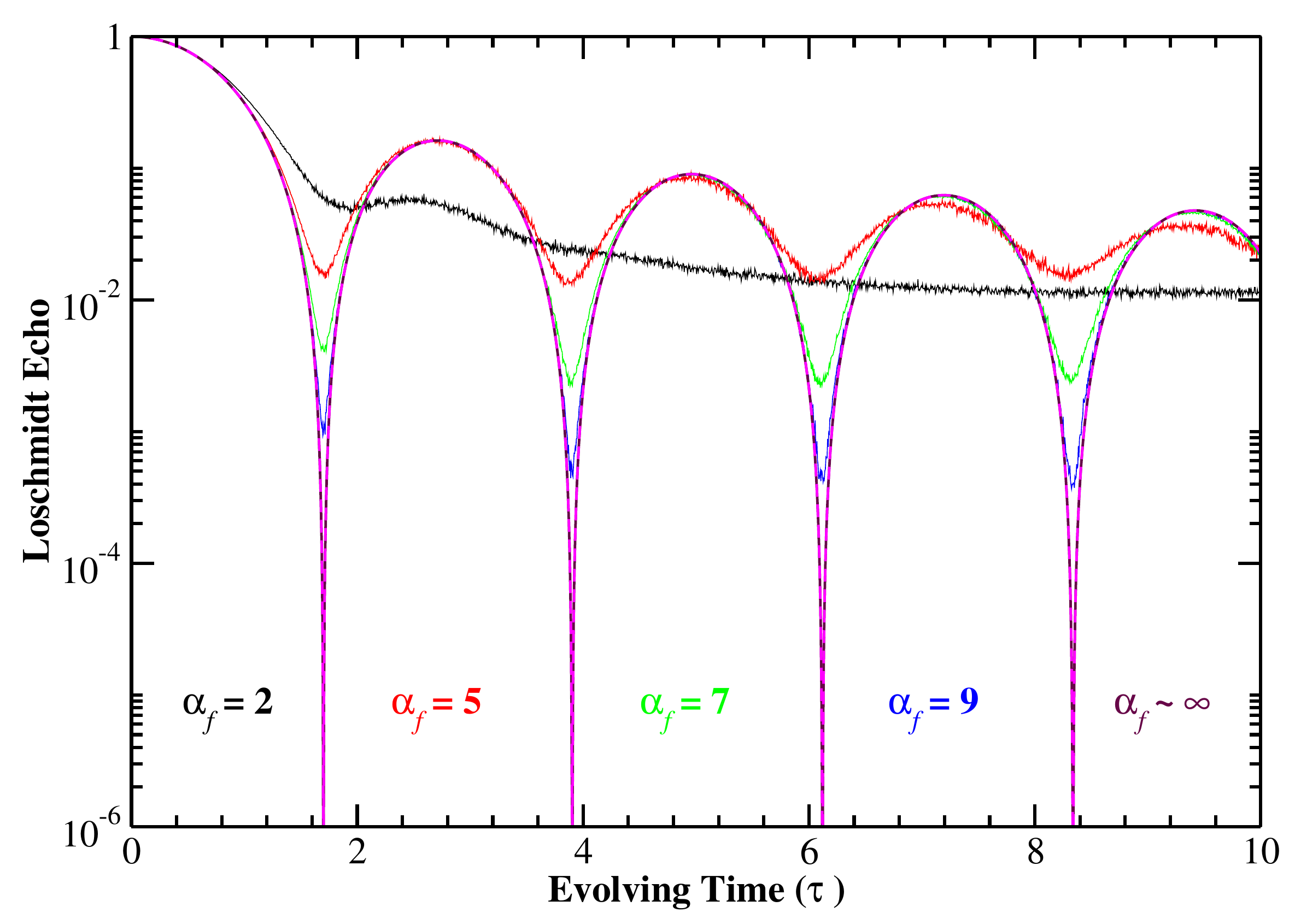}
\par\end{centering}
\caption{(Color online) Log-linear scale: The time-evolution of Loschmidt echo
for various quenched modulation correlation exponents, $\alpha_{f}$
with a system of size $N=512$. The initial state is fixed to be the
ground state of the prequenched Hamiltonian with zero diagonal potential.
The magenta dashed curve corresponding to the analytical result at
$\alpha_{f}=\infty$ in the thermodynamic limit.\label{fig:LEvsCorrelation-ei0ef-L512}}
\end{figure}
We now concentrate on the quench dynamics of the correlated disordered
system where the quench is characterized by an abrupt change of the strength
of spatial correlations in the diagonal random potential. Initially, the system is considered in a state $\ket{\Psi(\alpha_{i})}$, which is the eigenstate of the Hamiltonian $\hat{\mathcal{H}}(\alpha_{i})$
of prequenched correlation strength $\alpha_{i}$ at time $\tau=0$ and $\ket{\Psi(\alpha_{i},\alpha_{f},\tau)}$ be the time-evolving state after performing an abrupt quench dynamics to the final state of Hamiltonian $\hat{\mathcal{H}}(\alpha_{f})$. The Loschmidt echo takes the modified form as:
\begin{equation}
\mathcal{L}(\alpha_{i},\alpha_{f},\tau)=\left|\braket{\Psi(\alpha_{i})}{\Psi(\alpha_{i},\alpha_{f},\tau)}\right|^{2},\label{eq:LoschmidtEcho-1}
\end{equation}
where $\alpha_{f}$ defines the strength of the postquench modulation
correlation at time $\tau$.
\subsection{The quench dynamics between prequenched pure and postquenched random Hamiltonians}

The main focus is to study the quench dynamics under correlated model
in different regimes \citep{Pires2019,Niaz2020}. In the case of $(\varepsilon(\alpha_{i})=0),$ the initial eigenstate
of the prequench Hamiltonian $\hat{\mathcal{H}}(\alpha_{i})$ is a
plane-wave state $\ket{\Psi(\alpha_{i})}=\ket k$ with eigenenergy
$E_{k}=2t\cos(ka)$, where, $a$ represents the lattice spacing. After applying a sudden quench process in the internal
correlations of the disorder potential, the corresponding Loschmidt amplitude can
be expressed as,
\begin{equation}
\mathcal{G}(\alpha_{f},\tau)=\braket{k|e^{-i\tau\hat{\mathcal{H}}(\alpha_{f})}}k.\label{eq:LAmplitude-ei0}
\end{equation}
When an initial extended state is quenched into a strongly correlated
regime $(\alpha_{f}=\infty)$. Then, all the eigenstates $\ket{\Psi_{m}(\alpha_{f})}$
of the postquench Hamiltonian are delocalized with eigenenergy $E_{m}=\sqrt{2}\cos\left(\frac{2\pi}{N}m+\phi_{1}\right)$.
In this scenario, the Loschmidt amplitude can be modified as:
\begin{align}
\mathcal{G}(\alpha_{f}=\infty,\tau) & =\sum_{m=1}^{N}\bra ke^{-i\tau\hat{\mathcal{H}}(\alpha_{f})}\ket{\Psi_{m}(\alpha_{f})}\braket{\Psi_{m}(\alpha_{f})}k,\nonumber \\
 & =\frac{1}{N}\sum_{m=1}^{N}e^{-i\sqrt{2}\tau\cos\left(\frac{2\pi}{N}m+\phi_{1}\right)}.\label{eq:amplitude}
\end{align}
In the range of large system size, the phase $\varphi=(\frac{2\pi}{N}m+\phi_{1})$ is randomly distributed between $-\pi$ and $\pi$.
Therefore, we may rewrite the expression Eq.~(\ref{eq:amplitude})
as:
\begin{align}
\mathcal{G}(\alpha_{f}=\infty,\tau) & =\frac{1}{2\pi}\int_{-\pi}^{\pi}d\varphi e^{-i\sqrt{2}\tau\cos\left(\varphi\right)},\nonumber \\
 & =J_{0}(\sqrt{2}\tau),\label{eq:LAmplitude-ei0-aInf}
\end{align}
where $J_{0}(x_{s})$, is the first kind zero-order Bessel function, has a series of zeros $x_{s}$, with $s\in\mathbb{N}$. The analytical expression of the Loschmidt echo is given by:
\begin{equation}
\mathcal{L}(\alpha_{f}=\infty,\tau)=\left|J_{0}(\sqrt{2}\tau)\right|^{2}.\label{eq:LE-ei0-alpha2Inf}
\end{equation}
From this expression, it is clear that the Loschmidt echo has a series
of zeros at critical times $\tau^{*}=x_{s}/\sqrt{2}$, with $s$ set of positive roots. In small $s$ limit, the roots of $J_{0}(x)$ can be computed approximately
by Stokes\textquoteright s approximation \citep{Bowman1958},
\begin{equation}
x_{s}=\frac{\beta}{4}\left(1+\frac{2}{\beta^{2}}-\frac{62}{3\beta^{4}}+\frac{7558}{15\beta^{6}}\right),\quad\beta=\pi(4s-1).
\end{equation}
The occurrence of zeros in the Loschmidt echo indicates the localization transitions, referred as the dynamical phase transitions. It is worthwhile to mention that the extended time-evolved states of the postquench Hamiltonian with correlated disorder (infinite correlation exponent) are entirely different from those of conventional eigenstate (plane-wave) of the prequench pure system Hamiltonian. As a consequence, the Loschmidt amplitude —scalar product of plane-wave and extended time-evolved states— vanishes at critical times, signaling dynamical phase transitions. 

\begin{figure}[ht!]
\begin{centering}
\includegraphics[scale=0.42]{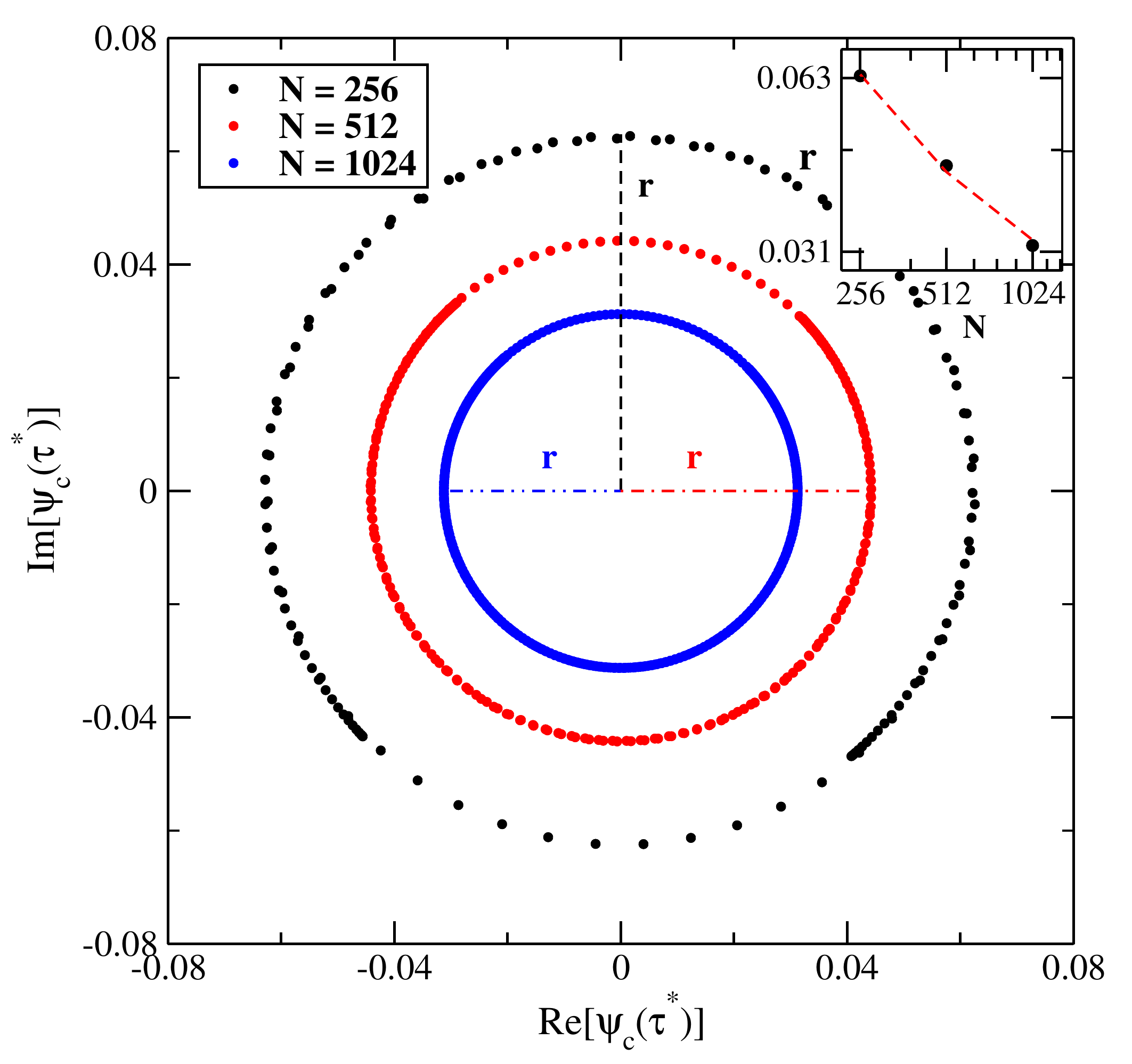}
\par\end{centering}
\caption{(Color online) The complex time-evolved state $\psi_{c}$, of the postquench Hamiltonian with modulation correlation exponent $\alpha_{f}\sim\infty$, for the system of sizes $N=256$ (black curve), $N=512$ (blue curve), and $N=1024$ (red curve) at critical time $\tau^{*}=3.9033$. The $\psi_{c}(\tau^{*})$ elements form a circular curves with center at the origin in the complex plane and \textcolor{black}{$r=0.063$} (black line), \textcolor{red}{$r=0.044$} (red line), and \textcolor{blue}{$r=0.032$} (blue line) are their corresponding radii. Inset: The radius of the same data as a function of system sizes in log-log scale. The data very well fitted with a curve, $y=a + b/x$ (red dashed curve). \label{fig:Eigenstate-Correlated}}
\end{figure}

Figure~\ref{fig:LEvsCorrelation-ei0ef-L512} illustrates the time evolution of Loschmidt echo when an initial
pure ground state ($\varepsilon(\alpha_{i})=0$) is quenched into
a correlated disordered regime. Here, the numerical calculations are carried out for the system of size $N=512$ and the sample average is taken over $1024$ realizations of disorder. However, the larger
$\alpha_{f}$ will smoother the random profile, and hence the Loschmidt echo due to the absence of the effective disorder. We demonstrate that
the Loschmidt echo tends to zero for $\alpha_{f}<1$ after some time
interval for a given realization of the disordered potential. Typically
when a pure state of system is quenched into an extended time-evolved
state, one may expect unit Loschmidt echo, as the prequench and postquench
states both are plane wave. On contrary, in the limit $\alpha_{f}\approx\infty$,
the Loschmidt echo exhibits singularities in time scale, verified
by the analytical result (magenta dashed curve) obtained in the thermodynamic
limit. This anomalous singular trend of the Loschmidt echo characterizes
the DQPTs in the quantum quenched system.

In order to know the origin of the anomalous dynamical phase transition, we calculate the eigenstates of the time-evolved state of the postquench Hamiltonian with infinite modulation correlation exponent. It is obvious that the eigenstates of a perfectly pure crystal are translationally invariant with probability amplitudes extending to all lattice sites. These extended states are explained by plane waves, which are the corresponding eigenstates of the system's Hamiltonian with energy spectrum $E_{k}=2t\cos ka.$ These eigenstates are
\begin{equation}
\ket{k}=\frac{1}{\sqrt{N}}\sum_{n=1}^{N}e^{ikan}\hat{c}_{n}^{\dagger}\ket 0,
\end{equation}
where $k$, denotes the wave vector lies in the first Brillouin zone
with $k\in(-\frac{\pi}{a},\frac{\pi}{a}]$. In Fig.~\ref{fig:Eigenstate-Correlated}, we manifest the distribution
of the complex time-evolved eigenstate elements for $\alpha_{f}=1000$ with various system sizes at critical time $\tau^{*}=x_{2}/\sqrt{2}=3.9033$.
At infinite disorder correlation, the eigenstates are perfectly ordered
(extended), however, are entirely different from those of conventional
eigenstate. The mean of the plane wave elements are exactly equal
to $\sqrt{N},$ whereas the average value of the time-evolved state
elements approaches to zero. Importantly, the weight of the positive and negative value of the complex state elements are approximately equal at critical times, resulting in vanishing overlap between
plane wave and its time-evolved state. In other words, the Loschmidt
amplitude turns out to be,
\begin{align}
\mathcal{G}(\alpha_{f}=\infty,\tau^{*}) & =\braket k{\psi_{c}(\tau^{*})},\nonumber \\
 & \approx0,\label{eq:abc}
\end{align}
in the limit of an infinite postquench disorder correlation strength. Here, $\ket{\psi_{c}(\tau^{*})}$, is the time-evolved state of the postquench Hamiltonian with diagonal correlated disorder at critical time. In the inset, we present the finite size scaling of the the time-evolved state at the critical time. It is noted that the radii varies as $a+b/x$ (red dashed line) obtained by fitting the data. This shows that radii of the time-evolved eigenstates elements curve approaches to zero in the thermodynamic limit.

\begin{figure}[ht!]
\begin{centering}
\includegraphics[scale=0.36]{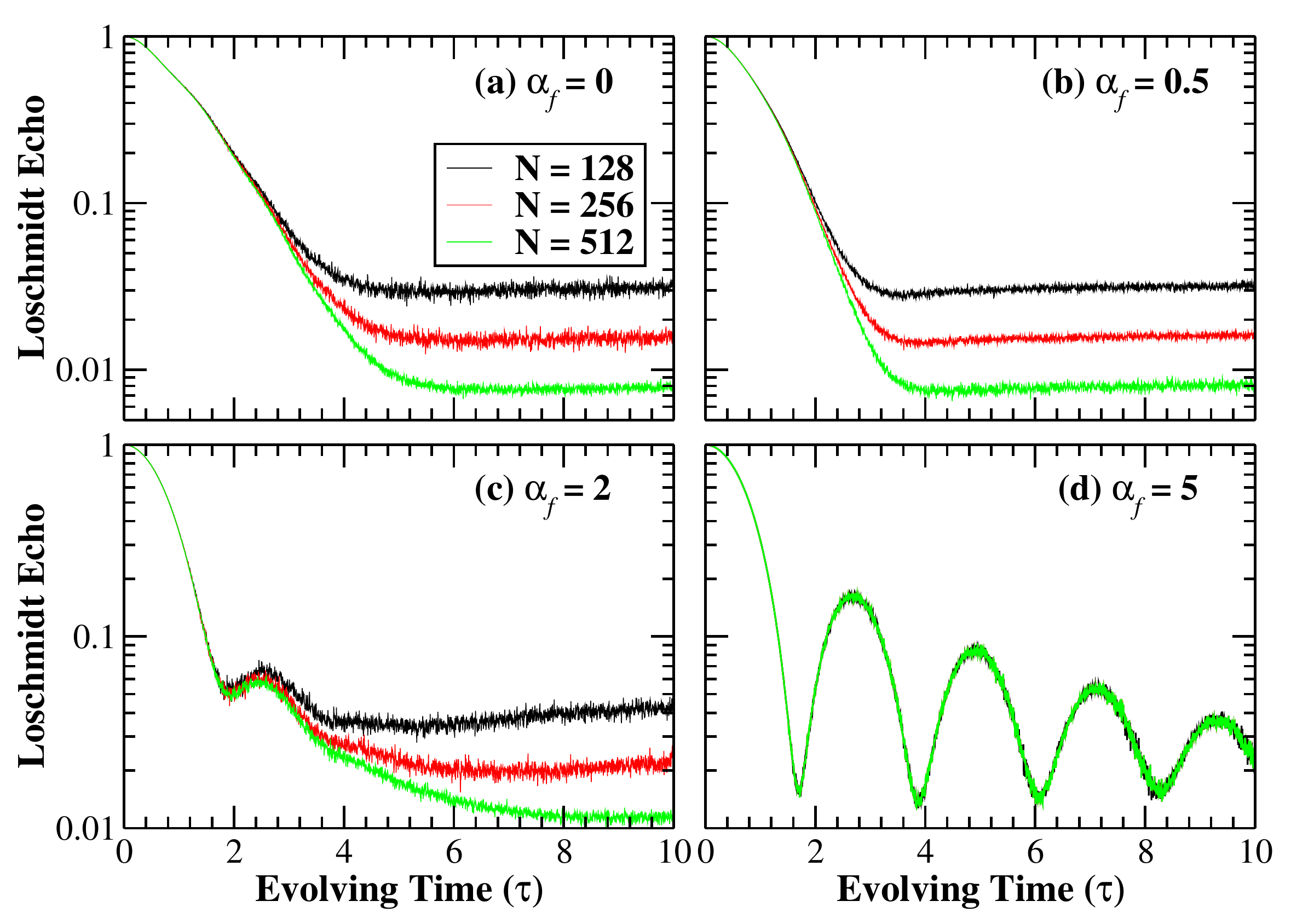}
\par\end{centering}
\caption{(Color online) Log-linear scale: The time-evolution of the Loschmidt
echo with $(\text{a})$ $\alpha_{f}=0$, $(\text{b})$ $\alpha_{f}=0.5,$
$(\text{c})$ $\alpha_{f}=2,$ and $(\text{d})$ $\alpha_{f}=5$ for different system sizes $N=128,\,256,\,\,\text{and}\,\,512$ with $2048$ realizations of disorder. The initial state is fixed to be the ground state of the prequenched Hamiltonian with $\varepsilon(\alpha_{i})=0$. While increasing the system size, the evolution of the Loschmidt echos decays monotonically for $\alpha_{f}<1$ after some time intervals. For $\alpha_{f}>1$, the Loschmidt echos decay either monotonically or periodically to zero.\label{fig:Echo-ei0-ef-Sizes}}
\end{figure}

\begin{figure}[ht!]
\begin{centering}
\includegraphics[scale=0.36]{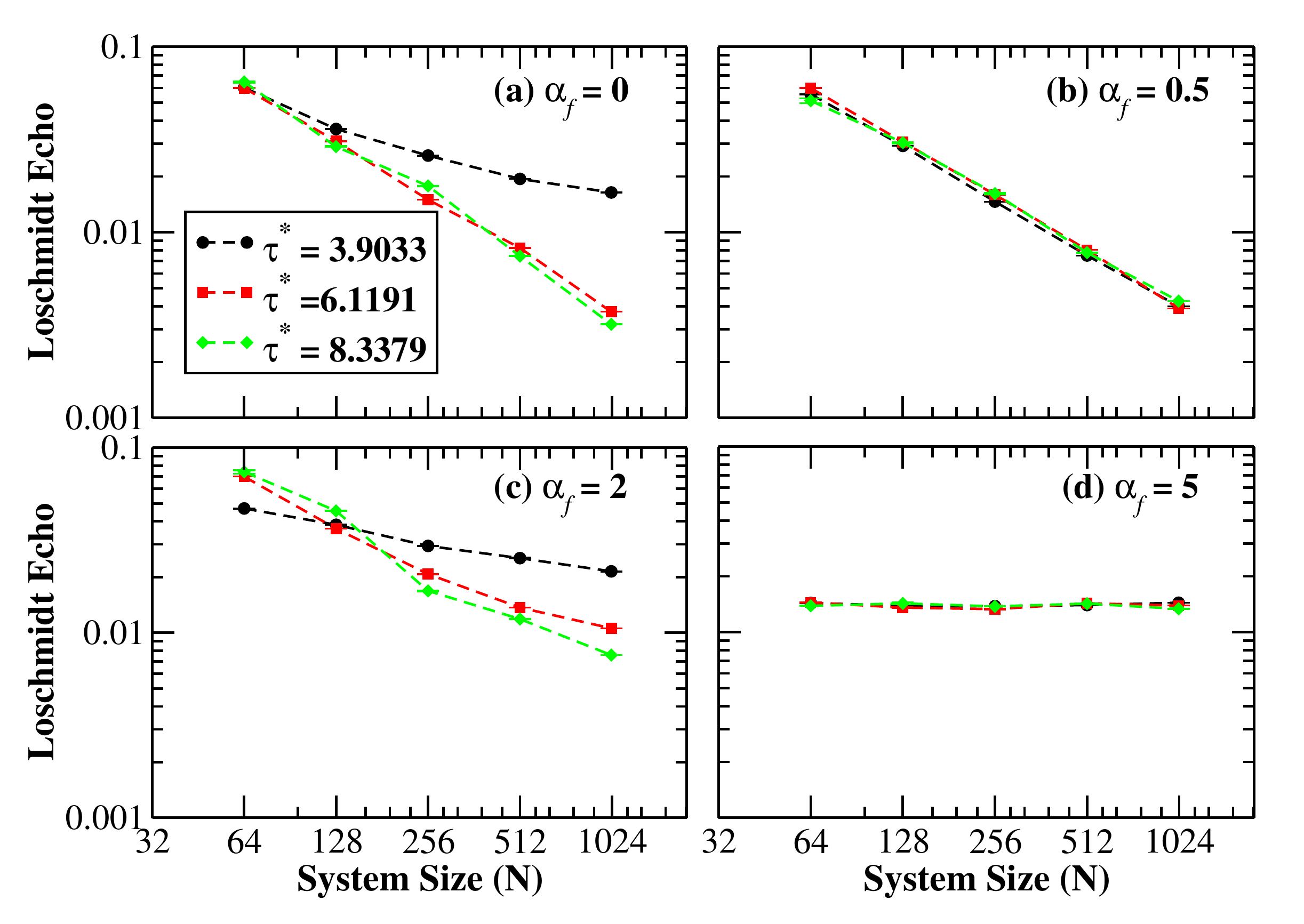}
\par\end{centering}
\caption{(Color online) Log-log scale: The time-evolution of the Loschmidt echo as a function of system sizes with $2048$ realizations of disorder of the data presented in Figure~\ref{fig:Echo-ei0-ef-Sizes} at critical times $\tau^{*}$$ = 3.9033$, $ 6.1191$, and $8.3379$.\label{fig:LEvsSizes-ei0-ef}}
\end{figure}

Furthermore, the quench dynamics under the correlated Anderson model for different system sizes are illustrated in Fig.~\ref{fig:Echo-ei0-ef-Sizes}. We find that the Loschmidt echo decreases with increasing system sizes after some time interval for finite correlations strength. However, the Loschmidt echo turned out to be size-independent in the strong correlation limit. This universal feature of the Loschmidt echo is expected to hold true as long as $\alpha_{a}$ is large enough. Moreover, we analyze the quench dynamics of the system for finite postquench correlation exponent as shown in Fig.~\ref{fig:LEvsSizes-ei0-ef}. In the localized regime, $\alpha_{f} \lesssim 1$, the Loschmidt echo decays as $y=e^{-\ln{x}}$ with system sizes as shown in Fig.~\ref{fig:LEvsSizes-ei0-ef} (a) and (b). However for $\alpha_{f} = 0$, we obtain a clear deviation from the curve, indicating the nonvanishing finite value of the Loschmidt echo in the thermodynamic limit, resulting from the shift in the curve with increasing size as shown in Fig.~\ref{fig:Echo-ei0-ef-Sizes}. On the other hand, $\alpha_{f} = 2$, the Loschmidt echo initially decrease to a minimum at critical point $\tau^{*}$$ = 3.9033$, then increases to a fixed point and the gradually decrease with time as shown in Fig.~\ref{fig:Echo-ei0-ef-Sizes} (c). Therefore, the echo will saturate to a fixed point with increasing system size as depicted in Fig.~\ref{fig:LEvsSizes-ei0-ef} (c). Moreover, the Loschmidt echo becomes universal in the strong disorder correlation limit as shown in Fig.~\ref{fig:LEvsSizes-ei0-ef} (d).
%
\subsection{The dynamics between prequench random and postquench pure Hamiltonians}

\begin{figure}
\begin{centering}
\includegraphics[scale=0.36]{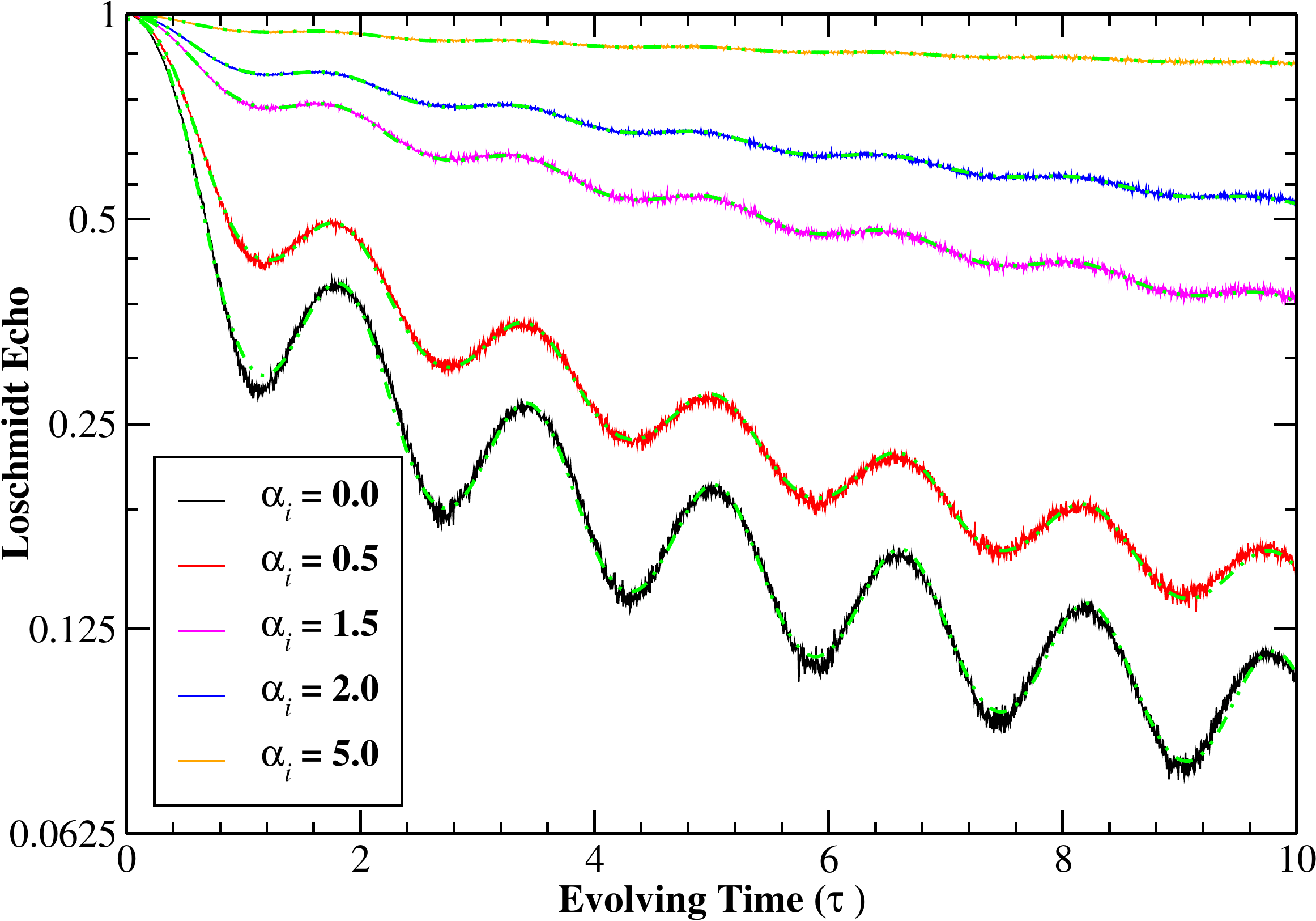}
\par\end{centering}
\caption{(Color online) Log-linear scale: The time-evolution of the Loschmidt
echo for various prequench correlation exponent $\alpha_{i}$ with
system size $N=512$ and sample averaged over $2048$ realizations of disorder.
The Loschmidt echoes are well fitted (green dashed-dotted curves) by Eq.~(\ref{eq:Fitting1})
for finite correlations of the disorder potential.\label{fig:LEvsCorrelation-ef0ei-L512}}
\end{figure}

Turning to the case where an initially prepared ground state of the
prequench Hamiltonian with diagonal correlated disorder is quenched
into an extended time-evolved state of the postquench Hamiltonian
with $\varepsilon(\alpha_{f})=0$. In the strongly localized regime
($\varepsilon(\alpha_{i})\rightarrow\infty$), one can analytically
obtain the evolution of Loschmidt echo, $\mathcal{L}(\tau)=\left|J_{0}(2\tau)\right|^{2}$
in the thermodynamic limit which is in excellent agreement with the
results as reported in the literature \citep{Yang2017,Xu2021}. Hence,
by construction, the ensemble average of the on-site energies are zero
and the local variance ---amplitude of random potential--- is site
independent and equal to unity \citep{Pires2019}. Figure~\ref{fig:LEvsCorrelation-ef0ei-L512}
shows the Loschmidt echo for various correlation exponent of the prequench
Hamiltonian. One can observe an oscillating decay of the Loschmidt
echos with evolving time which are very well fitted to the scaling function,
\begin{align}
\mathcal{L}(\tau) & =a_{0}e^{-a_{1}\tau}+\frac{a_{2}}{\tau+a_{3}}e^{-a_{4}\sin^{2}(a_{5}\tau-a_{6})},\label{eq:Fitting1}
\end{align}
where $a_{0}$, $a_{1},...,a_{6}$ are the fitting parameters. The first term in Eq.~(\ref{eq:Fitting1}) is dominant initially, where the Loschmidt echo decays exponentially for short interval of time, and then decay oscillatory after some time interval.

The Loschmidt echo initially reduces to a minimum value and then starts
to decay oscillatory with time interval. Importantly, for a fixed finite
system, the Loschmidt echo increases with correlations, tending to unity
in the limit $\alpha_{i}\rightarrow\infty$, where an overarching
sinusoidal structure begins to develop in the disorder configuration.
In this case, the system displays no signature of dynamical phase transition.
%
\subsubsection{Delocalization Transition}

\begin{figure}
\begin{centering}
\includegraphics[scale=0.36]{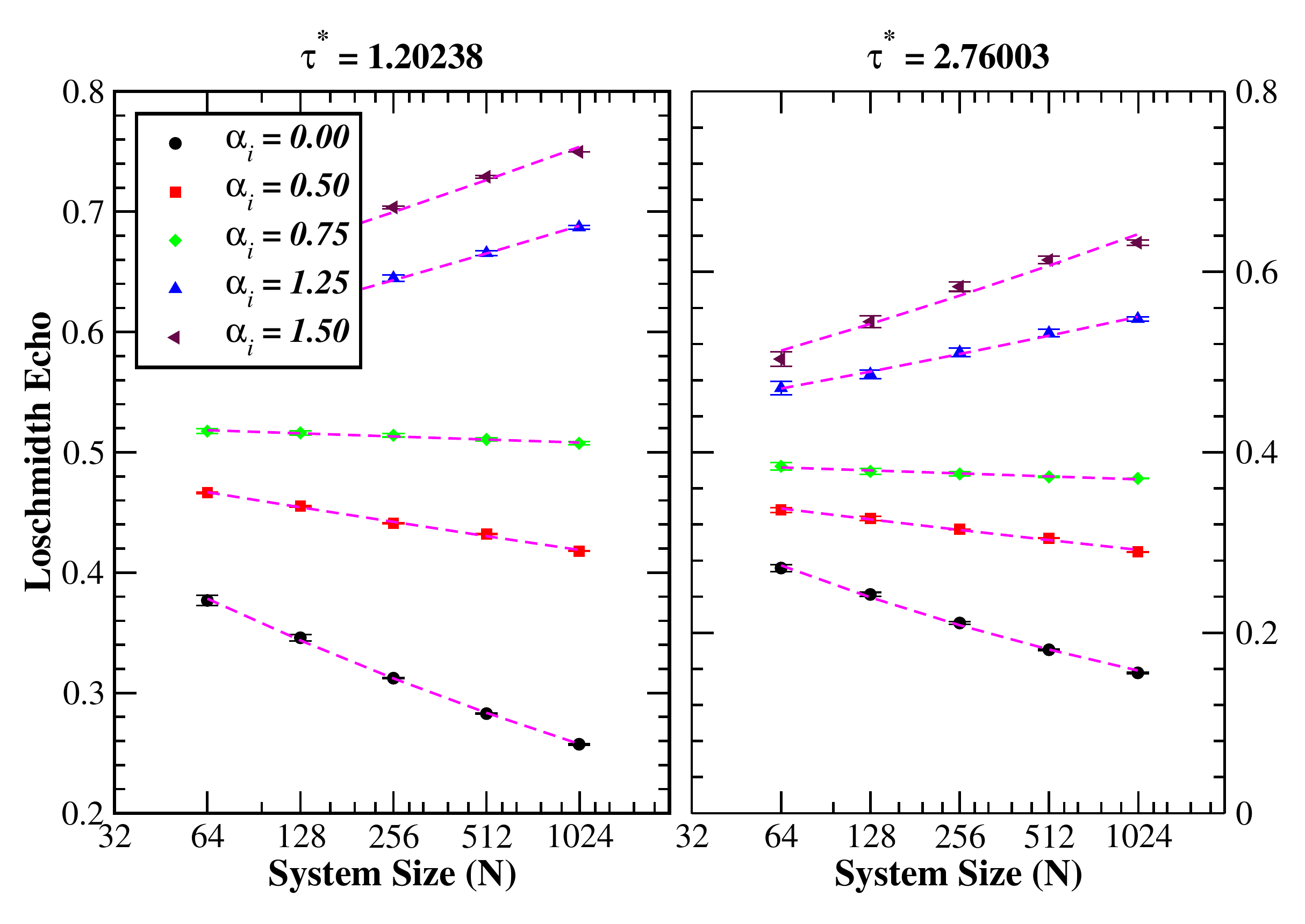}
\par\end{centering}
\caption{(Color online) Linear-log scale: Scaling of the Loschmidt echo for
various prequench correlation exponent $\alpha_{i}$ at critical evolving time
$\tau^{*}=1.20238$ (left panel) and $\tau^{*}=2.76003$ (right panel) and sample averaged over $2048$ realizations of disorder. The initial state is fixed to be the ground state of the prequench Hamiltonian with correlated potential. Loschmidt echos are well fitted (magenta curves) by Eq.~(\ref{eq:ScalingFunction}) for finite
correlations. The statistical error  (symbols with error bars) can be estimated by the standard deviations of the Loschmidt echo with the different system sizes.  The numerical data are linearized by using linear-log scale.
\label{fig:LEvsL-ef0ei}}
\end{figure}

Another important aspect of the quench dynamics concerns the size
scaling of the Loschmidt echo of the system. It turns
out to be a exponential decaying function of the system's size for
$\alpha_{i}<1$ at fixed evolving times as illustrated in Fig.~\ref{fig:LEvsL-ef0ei}.
Intuitively, it approaches to zero after some time interval in the range of
thermodynamic limit. Most importantly, the Loschmidt echo becomes
size-independent at the transition point $(\alpha_{i}\sim1)$. For
$\alpha_{i}>1$, however, the Loschmidt echo appears to grow exponentially
with system's sizes for $\alpha_{i}>1$, and tends to unity in the
thermodynamic limit. Moreover, the Loschmidt echos are very well fitted
by,
\begin{equation}
\mathcal{L}(\tau)=\begin{cases}
ae^{-b \tau} & \alpha_{i}<1,\\
a & \alpha_{i}\sim1,\\
ae^{b \tau} & \alpha_{i}>1,
\end{cases}\label{eq:ScalingFunction}
\end{equation}
where\textbf{ $a$ }and $b$ are positive real constant. Expression~(\ref{eq:ScalingFunction})
shows that the scaling function decays for $\alpha_{i}<1$, remains constant for $\alpha_{i}\sim1,$ and grows for $\alpha_{i}>1$, corresponding to the localized, critical and extended regime of the system, respectively.
Numerical studies have remarked on the smoothening of the disorder
amplitude with increased system size \citep{Russ2001,Petersen2013}.
However, we argue that this smoothing of the potential landscape happens
for $\alpha_{i}>1$.\textbf{ }On the contrary, one recovers the Anderson
model with uncorrelated disorder for $\alpha_{i}<1$\textbf{ }with
increasing system size. We assign this structure to be one of the
reasons for the emergence of delocalization transition in the system.
Further, using the generalized Thouless formula \citep{Izrailev1999},
the localization length $\xi$ of the correlated Anderson model for
$\alpha\lesssim1$ can be analytically calculated as \citep{Pires2019,Niaz2020},
\begin{equation}
\xi(E)=\frac{8}{(1-\alpha)\sigma_{\varepsilon}^{2}}(1-\frac{E^{2}}{4})\left[\frac{2}{\pi}\arccos(\frac{E}{2})\right]^{\alpha},\label{eq:LLMLmodel}
\end{equation}
in the thermodynamic limit at energy $E$. This result has been numerically verified
via calculating the localization length from the scaling
of the conductance \citep{Pires2019} and the kernel polynomial method
\citep{Niaz2020}. It is evident that the localization length diverges
as $\alpha\rightarrow1$ for any arbitrary value of the band energy, signaling
the existence of delocalization transition. Our results support the
idea that the delocalization transition occurs at $\alpha\sim1$,
in the thermodynamic limit. It turns out that the Loschmidt echo can
also be employed as a theoretical technique for the investigation of delocalization
phase transition in the correlated Anderson model.

\subsection{The quench dynamics between two independent random Hamiltonians}

\begin{figure}[ht!]
\begin{centering}
\includegraphics[scale=0.36]{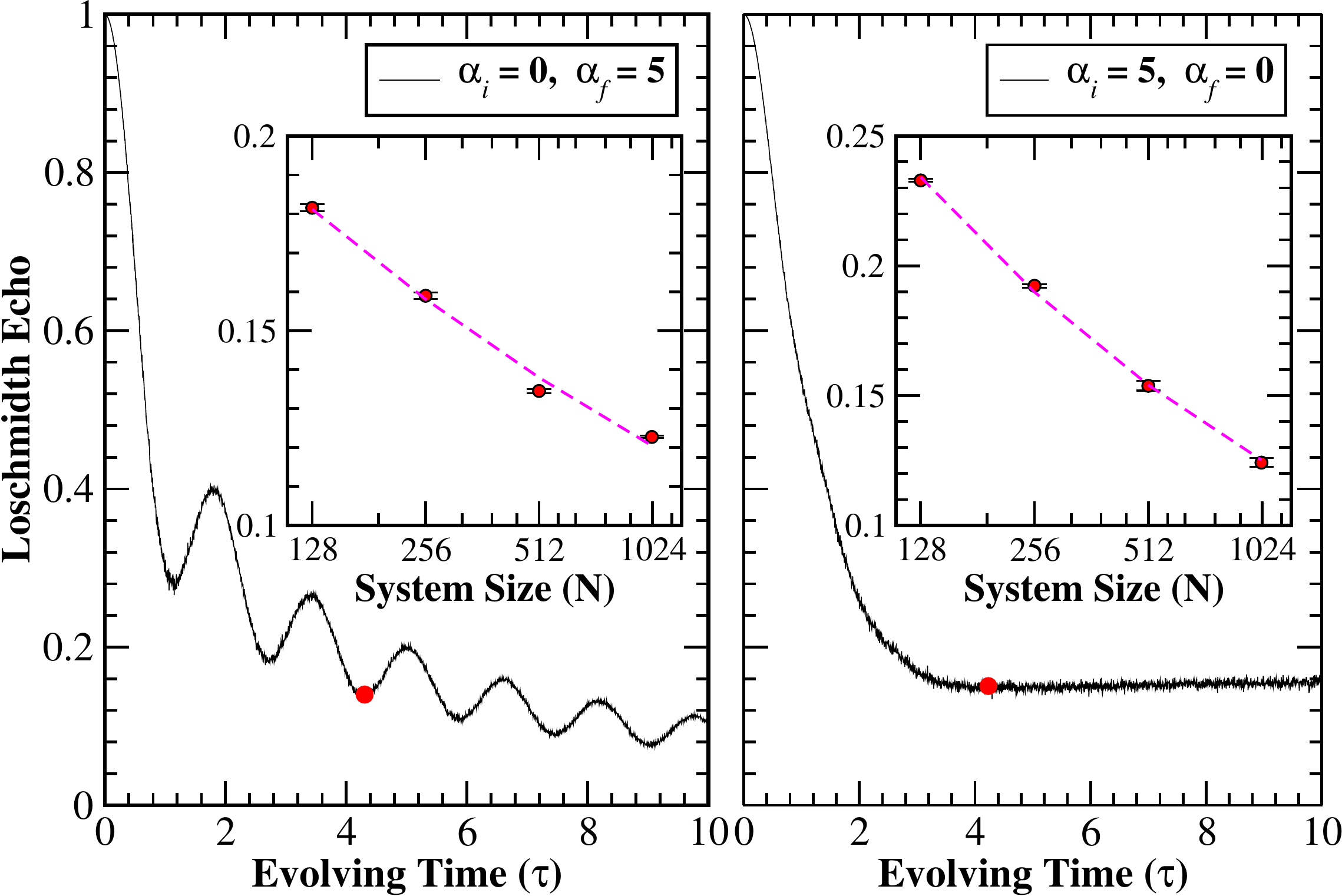}
\par\end{centering}
\caption{(Color online) The time-evolution of Loschmidt echo for $\alpha_{i}=0$
and $\alpha_{f}=5$ (left panel) and $\alpha_{i}=5$ and $\alpha_{f}=0$
(right panel) with system size $N=512$, averaging over $2048$ samples. Insets: Finite size scaling
of the Loschmidt echos of the corresponding fixed critical evolving time $\tau^{*}=4.3269$
(red point). The Loschmidt echos are well fitted (magenta dashed curves)
by an exponential decaying function $y=ae^{-bx},$ where $a$ and
$b$ are the fitting parameters. The statistical error  (symbols with error bars) can be estimated by the standard deviations of the Loschmidt echo with the different system sizes. The data are linearized by using
linear-log scale in insets.\label{fig:LEvsC-efei}}
\end{figure}

Further, exploring the quantum quench analysis for the scenario where
an initial ground state of the correlated disorder system is quenched
into a time-evolved state of the system with diagonal correlated disorder
potential. The quench dynamics between two independent random Hamiltonians
with $\alpha_{i}=0$ and $\alpha_{f}=5$, leading to an oscillating
decay of the Loschmidt echo after some interval of time as depicted
in Fig.~\ref{fig:LEvsC-efei} (left panel). The result bears a striking
resemblance to the data presented in Fig.~\ref{fig:LEvsCorrelation-ef0ei-L512}
where a reference state $(\alpha_{i}=0)$ is quenched into a time-evolved
extended state of the system with zero diagonal potential $(\varepsilon(\alpha_{f})=0)$.
Indeed, one would expect similar results as the time-evolved state
where both the cases are extended. However, one may get a small deviation
of the Loschmidt echo for a finite system when $\alpha_{f}$ approaching
to critical region. In the inset we show that the Loschmidt echo exponentially
decays to zero in the thermodynamic limit. For the case where $\alpha_{i}=5$
and $\alpha_{f}=0$, the Loschmidt echo monotonically decays to a
finite value after some interval of time as shown in Fig.~\ref{fig:LEvsC-efei}
(right panel). However, the system displays DQPTs, characterized by
the vanishing value of Loschmidt echo in the thermodynamic limit (inset).
Further, the favorable scaling features of Loschmidt echo become vital,
as they make it possible to predict the nature of the correlated Anderson model.

Typically, Loschmidt echo decays from unity, oscillates with same
frequency and damping amplitude after some interval of time, if an initially
extended state is quenched into a strongly localized regime
\citep{Yang2017,Yin2018,Xu2021}. However, the quench dynamics under
correlated Anderson model reveals that the Loschmidt echo qualitatively
shows similar decaying behavior at different critical times, if the
initial extended state is quenched into a strongly correlated regime.

A fascinating road map of research is the mutual interaction between correlations in the hopping integrals and on-site energies. As shown, correlations in the on-site disorder potential may trigger the dynamical phase
transitions depending on the correlation controlling parameter and
the quenching process. An intriguing follow-up of our present work
would be the investigation of dynamical phase transitions in the model
with power-law correlation hopping integral.

\section{Conclusions}

We studied the nonequilibrium dynamics of the 1D non-interacting correlated Anderson model, where the quench dynamics are induced by an abrupt change in the strength of disorder correlations. The system displayed an anomalous dynamical phase transition when an initial pure ground state is quenched into a strongly correlated disorder regime. In this limit, the disorder correlations induce cusp-like singularities in the Loschmidt echo at critical times, which are confirmed by analytical calculations in the thermodynamic limit. In other words, the overlap between the plane wave and its time-evolved delocalized state exhibited a series of zeros periodically with critical times, reflecting the anomalous DQPTs. Furthermore, the system showed universal size scaling behavior in the strong disorder correlations. On the contrary, the Loschmidt echo decays monotonically for the postquench white-noise potential (time-evolved localized state). Moreover, the Loschmidt echo turned out to be size-dependent for the Anderson-like potential.

The dynamics between the prequench random and postquench pure Hamiltonians have also been investigated. It is pointed out that the Loschmidt echos monotonically decay before a finite time and then undergoes an oscillatory decay with time, an initial localized state (Anderson-like disorder) into a time-evolved extended regime (zero on-site potential). The Loschmidt echo quantitatively increases with increasing disorder correlations and approaches unity in the infinite correlation limit.
However, the decay of Loschmidt echo is enhanced (suppressed) by increasing the system size for an initially localized (delocalized) regime. As a consequence, the system exhibited the DQPTs for an initially localized state with Anderson-like disorder in the thermodynamic limit. Whereas, the Loschmidt echos turned out to be unity for an initially delocalized state $(\alpha_{i}>1)$ in the thermodynamic limit. Furthermore, this scaling behavior of Loschmidt echo can also be mapped with the identification of a correlation-induced delocalization phase transition in the correlated Anderson model.

\section*{Author Contribution Statement}
N.A.K. developed the theoretical formalism and performed the analytic and numerical simulations, as verified by M.J., P.W., and G.X. N.A.K. wrote the original manuscript, reviewed by the P.W., and M.J. G.X. funded and supervised the project. All authors discussed the results and contributed to the final manuscript.
\begin{acknowledgments}
N.A.K. and M.J. acknowledge the postdoctoral fellowship supported
by Zhejiang Normal University under Grants No. ZC304022980 and No.
ZC304022918, respectively. G.X. acknowledges support from the NSFC
under Grants No. 11835011 and No. 12174346. 
\end{acknowledgments}

\section*{Disclosures}
The authors declare no conflicts of interest.

\section*{Data Availability Statement}
The datasets used and/or analyzed during the current study are available from the corresponding author on reasonable request.

\bibliographystyle{apsrev4-2.bst}
\bibliography{cDQPT}

\end{document}